\documentclass[prb,citeautoscript,amsmath,twocolumn,amssymb,superscriptaddress]{revtex4-2}
\usepackage[utf8]{inputenc}
\usepackage{xspace}
\usepackage{physics}
\usepackage{graphicx}
\usepackage[dvipsnames]{xcolor}
\usepackage[separate-uncertainty=true,multi-part-units=single]{siunitx}
\usepackage[version=4]{mhchem}
\usepackage{xr-hyper}
\usepackage[colorlinks, citecolor={blue}, linkcolor={black}, urlcolor={blue}]{hyperref}

\newcommand{\figlbl}{Fig.}

\newcommand{\figtitle}[1]{\textbf{#1}\xspace} 

\newcommand{\panela}{\textbf{a}\xspace}
\newcommand{\panelb}{\textbf{b}\xspace}
\newcommand{\panelc}{\textbf{c}\xspace}
\newcommand{\paneld}{\textbf{d}\xspace}
\newcommand{\panele}{\textbf{e}\xspace}
\newcommand{\panelf}{\textbf{f}\xspace}
\newcommand{\panelg}{\textbf{g}\xspace}
\newcommand{\moire}{moir{\'e}\xspace}

\begin{document}
\title{Cascade of isospin phase transitions in Bernal bilayer graphene at zero magnetic field}

\author{Sergio C. de la Barrera}
\author{Samuel Aronson}
\author{Zhiren Zheng}
\affiliation{Department of Physics, Massachusetts Institute of Technology, Cambridge, MA, USA}
\author{Kenji Watanabe}
\affiliation{Research Center for Functional Materials, National Institute for Materials Science, Tsukuba, Japan}
\author{Takashi Taniguchi}
\affiliation{International Center for Material Nanoarchitectonics, National Institute for Materials Science, Tsukuba, Japan}
\author{Qiong Ma}
\affiliation{Department of Physics, Boston College, Chestnut Hill, MA, USA}
\author{Pablo Jarillo-Herrero}
\author{Raymond Ashoori}
\affiliation{Department of Physics, Massachusetts Institute of Technology, Cambridge, MA, USA}

\begin{abstract}

Emergent phenomena arising from the collective behavior of electrons is generally expected when Coulomb interactions dominate over the kinetic energy, as in delocalized quasiparticles in highly degenerate flat bands.
Bernal-stacked bilayer graphene intrinsically supports a pair of flat bands predicted to host a variety of spontaneous broken-symmetry states arising from van Hove singularities and a four-fold spin-valley (isospin) degeneracy \cite{min2008pseudospin,nandkishore2010quantum,lemonik2010spontaneous,weitz2010brokensymmetry,zhang2011spontaneous,jung2011lattice}.
Here, we show that ultra-clean samples of bilayer graphene display a cascade of symmetry-broken states with spontaneous and spin and valley ordering at zero magnetic field.
Using capacitive sensing in a dual-gated geometry, we tune the carrier density and electric displacement field independently to explore the phase space of transitions and probe the character of the isospin order.
Itinerant ferromagnetic states emerge near the conduction and valence band edges with complete spin and valley polarization and a high degree of displacement field tunability.
At larger hole densities, two-fold degenerate quantum oscillations manifest in an additional broken symmetry state that is enhanced by the application of an in-plane magnetic field.
Both types of symmetry-broken states display enhanced layer polarization at low temperatures, suggesting a coupling to the layer pseudospin degree of freedom in the electronic wavefunctions \cite{min2008pseudospin,jung2011lattice}.
Notably, the zero-field spontaneous symmetry breaking reported here emerges in the absence of a \moire superlattice and is intrinsic to natural graphene bilayers.
Thus, we demonstrate that the tunable bands of bilayer graphene represent a related, but distinct approach to produce flat band collective behavior, complementary to engineered \moire structures.

\end{abstract}

\maketitle

Large electronic densities of states enhance the effects of interactions between electrons and give rise to collective electronic states such as Stoner ferromagnetism and BCS superconductivity.
In the well-known example of quantum Hall ferromagnetism, a perpendicular magnetic field transforms a two-dimensional electronic system into a series of macroscopically degenerate flat bands that spontaneously spin polarize at half-filling.
Magic-angle twisted bilayers and trilayers of graphene develop flat bands at zero magnetic field due to the influence of a \moire superlattice potential.
Following discoveries of correlated insulators, superconductivity, ferromagnetism, and a host of other collective phenomena at partial filling in \moire structures \cite{cao2018correlated,cao2018unconventional,sharpe2019emergent,yankowitz2019tuning,lu2019superconductors,serlin2020intrinsic,zondiner2020cascade,wong2020cascade,park2021tunable,hao2021electricfield}, attention has turned to material systems with enhanced densities of states arising from flat bands at zero magnetic field.
Few-layer graphene with rhombohedral (R)-stacking also generates highly degenerate flat bands in zero field but without the requirement of a \moire superlattice \cite{castro2008lowdensity,koshino2009trigonal,zhang2010bandstructure,zhang2011spontaneous,shi2020electronic}.
Bernal-stacked bilayer graphene, equivalent to R-stacking, is the simplest member of this family of structures, however, to date reports of symmetry-broken states in bilayer graphene have generally required large magnetic fields \cite{zhao2010symmetrybreaking,maher2013evidence,ki2014observation,zibrov2017tunableinteracting} or involved subtle gapped states at charge neutrality \cite{feldman2009brokensymmetry,mayorov2011interaction,freitag2012spontaneously,bao2012evidence,nam2018family,geisenhof2021quantum}.
Intriguingly, R-stacked (ABC) trilayer graphene was recently shown to exhibit spontaneous itinerant spin and valley polarization at zero magnetic field, displaying a series of phase transitions, hysteresis, anomalous Hall, and superconductivity at millikelvin temperatures in carefully prepared samples \cite{zhou2021halfandquarter,zhou2021superconductivity}.
R-stacking is crucial for these phenomena, however this stacking arrangement is not stable in trilayers, inhibiting experimental efforts.

\begin{figure}
    \centering
    \includegraphics[width=3.375in]{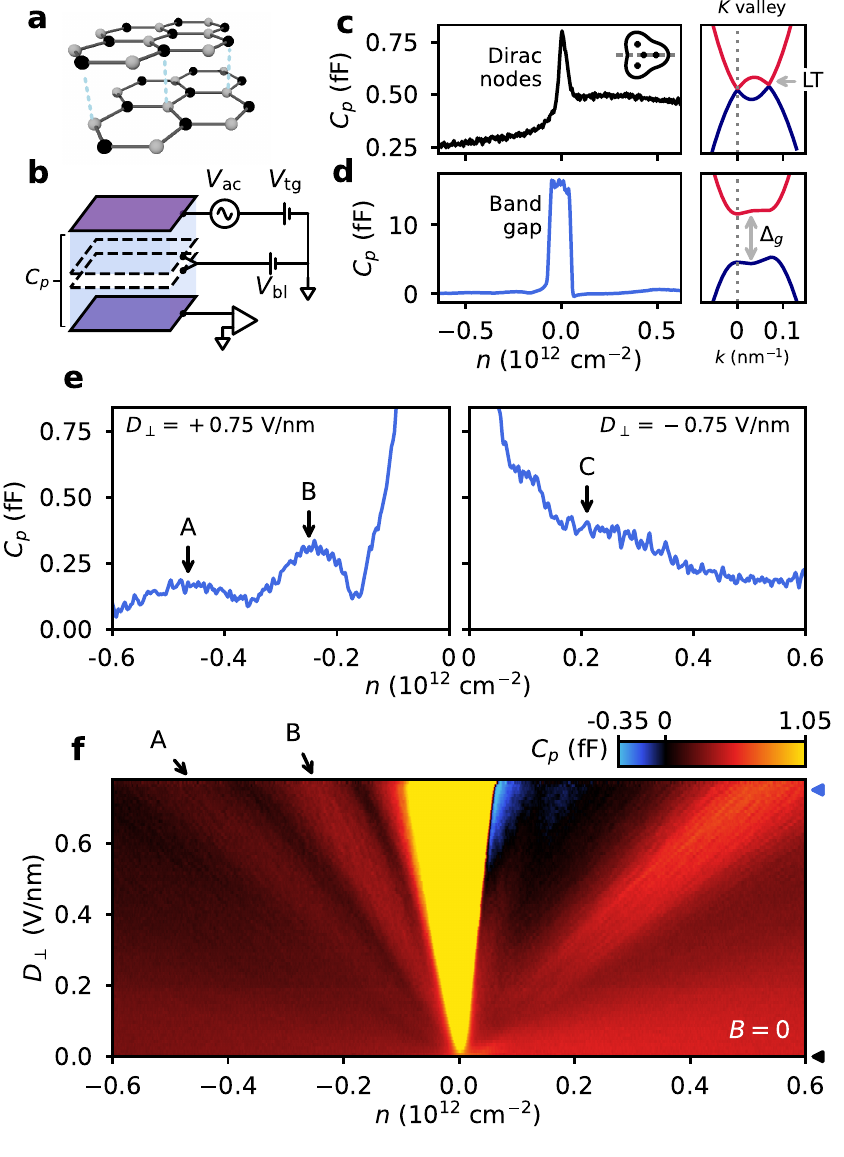}
    \caption{\figtitle{Flat bands in Bernal bilayer graphene.}
    \panela Bernal-stacked (AB) bilayer graphene structure.
    \panelb Schematic of penetration field capacitance $C_p$ measurement. Effective gate voltages are defined as $V_t = V_{tg} - V_{bl}$ and $V_b = -V_{bl}$, where $V_{bl}$ is the dc voltage applied to the bilayer graphene.
    \panelc Measured $C_p$ versus carrier density $n$ at zero electric displacement, $D_\perp = 0$. Right: corresponding band structure along $K\rightarrow\Gamma$ direction showing two of four Dirac nodes. Inset: Fermi contour showing effect of trigonal warping around $K$ and relative position of the Dirac nodes (points). The dashed line indicates the $k$ slice along which the bands are shown.
    \paneld Measured $C_p$ at $D_\perp = \SI{0.75}{V/nm}$ showing an incompressible peak in the band gap of the bilayer. Right: band structure with emerging flat bands driven by $D_\perp$-field.
    \panele Detail of trace shown in \panele showing two prominent oscillations on the hole side of the incompressible peak, labeled A and B, with corresponding trace from $D<0$ showing subtle oscillations on the electron side of charge neutrality, with label C indicating a broad shoulder region.
    \panelf $C_p$ map measured as a function of $n$ and $D_\perp$ at \SI{40}{mK}. Positions of the oscillations in \panele depend linearly on the magnitude of $D_\perp$.
    }
    \label{fig1}
\end{figure}

Bernal-stacked (AB) bilayer graphene on the other hand is stable in the R-stacking configuration (\figlbl\ref{fig1}\panela) and inherits a qualitatively similar band structure (\figlbl~\ref{fig1}\panelc-\paneld) to that of ABC trilayer graphene \cite{castro2008lowdensity,koshino2009trigonal,zhang2010bandstructure,zhang2011spontaneous}, with saddle points and flat bands occurring at lower energies (\SI{\sim 1}{meV}) compared to trilayers (\SI{\sim 10}{meV}).
As a result, correlation-driven phenomena may also manifest in AB bilayers in samples with sufficiently low charge disorder \cite{zheng2020unconventional}.

To explore this possibility, we probe the electronic compressibility of ultra-clean bilayer graphene samples at low carrier densities and in the presence of perpendicular displacement fields.
We fabricate dual-gated capacitor devices using graphite for the top and bottom gate electrode material and measure the penetration field capacitance $C_p$, related to the inverse compressibility $(\partial n/\partial\mu)^{-1}$, from the top gate to the bottom gate electrode (see \figlbl~\ref{fig1}\panelb, also Methods).
Depending on the compressibility of the bilayer graphene, $C_p$ varies between the geometric capacitance between the top and bottom gates (when the bilayer is incompressible or gapped) and zero (when the bilayer is infinitely compressible, as in a metal).

Using the top and bottom gate voltages (see \figlbl~\ref{fig1}\panelb), we vary the carrier density $n = (C_t V_t + C_b V_b)/e$, while holding the out-of-plane displacement field $D_\perp = (C_t V_t - C_b V_b)/2\epsilon$ fixed.
At $D_\perp=0$, the measured $C_p$ shows a jump in the charge compressibility at charge neutrality, with holes exhibiting higher compressibility compared to electrons (\figlbl~\ref{fig1}\panelc).
In the presence of a large, positive displacement field, the small peak at charge neutrality develops into a large, incompressible peak (\figlbl~\ref{fig1}\paneld), as expected for the tunable band gap in the band structure \cite{mccann2006asymmetry,zhang2009direct,hunt2017direct}.
A series of peaks appears on either side of charge neutrality (\figlbl~\ref{fig1}\panele) that do not match features in the single-particle electronic structure (see \figlbl.~\ref{fig1}\panelc-\paneld).
The non-interacting bands of bilayer graphene develop a divergence in the density of states on each side of the gap arising from a van Hove singularity and Lifshitz transition (LT in \figlbl~\ref{fig1}\panelc) at the saddle point within each band \cite{castro2008lowdensity,mccann2013electronic}.
However, two peaks appear on the hole side of the gap, labeled A and B in \figlbl~\ref{fig1}\panele, separated by two highly compressible minima.
By measuring these peaks as a function of $D_\perp$ (\figlbl~\ref{fig1}\panelf), we see that their location in electron density varies linearly with displacement field over a wide range and trends to zero as $D_\perp\rightarrow 0$.

\begin{figure}
    \centering
    \includegraphics[width=3.375in]{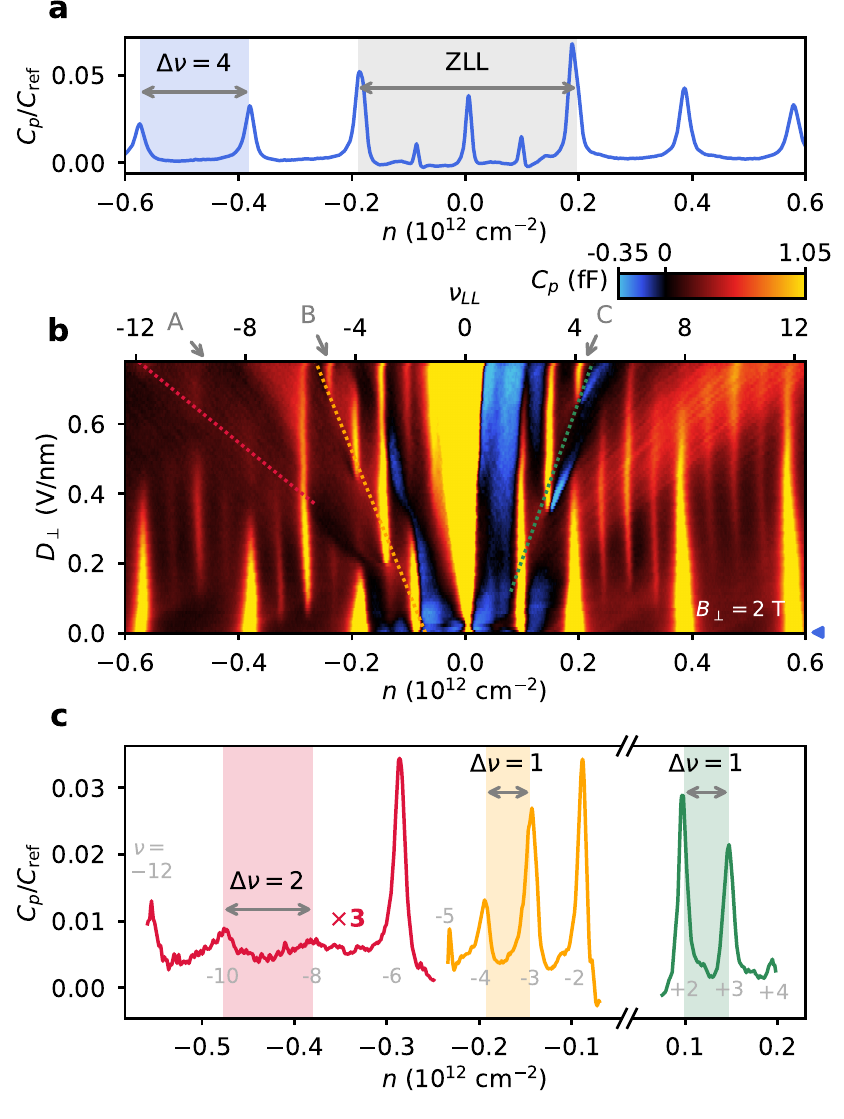}
    \caption{\figtitle{Degeneracy of broken-symmetry states in $B_\perp$.}
    \panela $C_p$ trace measured at $B_\perp=\SI{2}{T}$ and $D_\perp=0$ with $2\times$-degenerate incompressible peaks (separation between subsequent filling factors, $\Delta\nu$) in the zero-energy Landau level (ZLL) and $4\times$-degenerate peaks elsewhere.
    \panelb Map of $C_p$ measured at $B_\perp=\SI{2}{T}$ versus $n$ and $D_\perp$, with Landau level filling factors ($\nu_\text{LL}$) shown above.
    \panelc Detail of line traces taken along the matching dotted lines in \panelb, showing sharply different $\Delta\nu$ from $D_\perp=0$.
    }
    \label{fig2}
\end{figure}

To understand the origin of the $D_\perp$-tunable peaks in compressibility, we explore quantum oscillations in the presence of a small perpendicular magnetic field, $B_\perp$, with respect to the graphene bilayer.
In \figlbl.~\ref{fig2}\panelb, we show a map of $C_p$ measured at $B_\perp = \SI{2}{T}$.
A series of cyclotron gaps are evident in the $C_p$ map as narrow vertical features at integer filling factors, $\nu\equiv ne/hB$, appearing due to Landau level formation in the bilayer.
Near $D_\perp=0$, the density separation between cyclotron gaps is $\Delta n \approx \SI{9.7e10}{cm^{-2}}$ or $\Delta\nu=2$ from $\nu=-4$ to $\nu=+4$, and $\Delta n \approx \SI{1.9e11}{cm^{-2}}$ or $\Delta\nu=4$ for $|\nu|\geq 4$ (\figlbl.~\ref{fig2}\panela).
The set of states from $-4 > \nu > 4$ emerges from the zero-energy Landau level (ZLL), characterized by degenerate spin, valley, and Landau level orbital quantum numbers.
A density separation of $\Delta\nu=2$ or $\Delta\nu=1$ is expected in the ZLL due to Coulomb-driven exchange splitting, particularly in the presence of a layer-polarizing displacement field.
Outside of the ZLL and at small magnetic fields, exchange generally lifts the four-fold degeneracy of each Landau level gradually as displacement field is increased from zero.
However, at finite displacement field and within the boundaries of regions A and B observed at $B=0$ (between the $C_p$ minima at $B=0$), the Landau levels exhibit sharp transitions to $\Delta\nu=1$ within region B and $\Delta\nu=2$ within region A (\figlbl.~\ref{fig2}\panelb).
Starting at a finite $D_\perp$ that depends on density, cyclotron gaps emerge for $\nu=-3,-4,-5$ within region B and $\nu=-6,-8,-10$ within region A.
Interestingly, a region with $\Delta\nu=1$ also emerges for electrons (labeled C in \figlbl.~\ref{fig2}\panelb), though the parent state at zero field is not as evident as in regions A and B.

\begin{figure}
    \centering
    \includegraphics[width=3.375in]{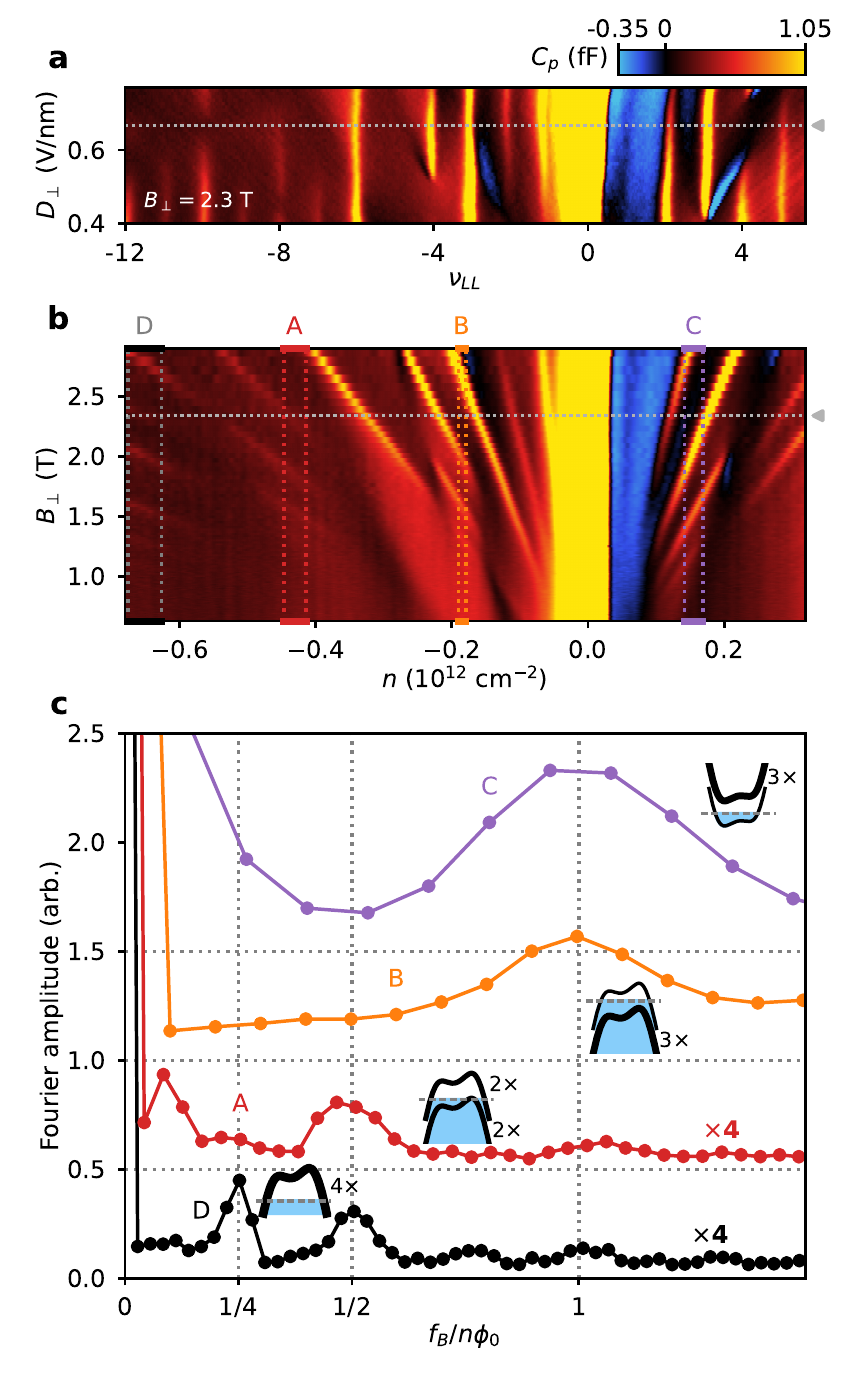}
    \caption{\figtitle{Fermi surface signatures in quantum oscillations of the broken-symmetry states.}
    \panela Density (filling factor, $\nu_\text{LL}$) versus $D_\perp$ map of $C_p$ measured at $B_\perp=\SI{2.3}{T}$.
    \panelb Measured $C_p$ versus $n$ and $B_\perp$ at fixed $D_\perp=\SI{0.67}{V/nm}$, along dashed line in \panela.
    \panelc Fast Fourier transform of magneto-oscillations from \panelb taken along $1/B_\perp$ and averaged over the range of densities indicated by the boxed regions in \panelb. The resulting frequencies are normalized by the total carrier density $n$ and magnetic flux quantum $\phi_0$ for cross-comparison.
    }
    \label{fig3}
\end{figure}

The observation of sharp transitions between Landau level spacing, $\Delta\nu$, at fixed $B_\perp$ suggests a sequence of phase transitions between phases with broken spin-valley (isospin) symmetry.
That is, the changes in Landau level degeneracy are driven by $D_\perp$-field induced spontaneous symmetry-breaking, even at relatively low magnetic fields, in contrast to the usual exchange-driven symmetry breaking generated by large magnetic fields.
At $B=0$ these transitions manifest as sharp minima in $C_p$, whereas the symmetry-broken regions occur within regions of reduced compressibility (as in the labeled peaks in \figlbl~\ref{fig1}\panele).

The assignment between $\Delta\nu$ and Fermi surface degeneracy within each region at fixed field can be further supported by investigating the magnetic field dependence of the symmetry-broken states as $B_\perp\rightarrow 0$.
In \figlbl.~\ref{fig3}\panelb, we show the field dependence of the Landau levels taken along a density cut at fixed $D_\perp$ (along the dashed line in \figlbl.~\ref{fig3}\panela).
The displacement field is fixed at $D_\perp=\SI{0.67}{V/nm}$ to sample a portion of each symmetry-broken region as well as the four-fold degenerate regions away from the symmetry-broken states.
Quantum oscillations with differing frequencies are observed in the resulting field dependence with clear boundaries near the edges of each symmetry-broken region (\figlbl.~\ref{fig3}\panelb).
Taking the Fourier transform along the field direction in $1/B_\perp$ for a range of densities in each region (labeled at the top of \figlbl.~\ref{fig3}\panelb) reveals the dominant oscillation frequency in each sampling window (\figlbl.~\ref{fig3}\panelc).

This oscillation frequency depends linearly on the total carrier density, $n$.
To compare frequencies over a range of densities, we normalize the oscillation frequency $f_B$ by the product of the total carrier density with the magnetic flux quantum, $n\phi_0$ with $\phi_0=h/e$, equivalent to the area of $k$-space enclosed by an orbit around the Fermi surface.
Using this normalization, Fourier transforms can be averaged over a window of densities (as indicated by the boxed regions in \figlbl.~\ref{fig3}\panelb) and plotted together (\figlbl.~\ref{fig3}\panelc).
At large hole densities (as in region D), the resulting Fourier transform displays a dominant peak at $f_B/n\phi_0 = 1/4$ (with additional harmonics), indicating that the carriers are divided evenly between four coexisting Fermi surfaces, as expected for a four-fold ($4\times$) degenerate band.
Within the intermediate density range at which $\Delta\nu=2$ Landau levels are shown at \SI{2}{T} in \figlbl.~\ref{fig2}\panelb-\panelc (region A), Fourier analysis reveals a peak at $f_B/n\phi_0 = 1/2$, or a reduction to two Fermi surfaces.
The peak at $1/2$ is consistent with the separation of Landau levels $\Delta\nu=2$ at fixed field, and further suggests that the system has spontaneously developed an exchange splitting that favors filling two of the four isospin flavors.
Notably, an additional low-frequency peak emerges in curve A that indicates some coexistence between Fermi surfaces with differing area, to be discussed shortly.
In regions B and C (for holes and electrons, respectively) peak Fourier amplitude occurs at $f_B/n\phi_0 = 1$.
Thus, only a single Fermi surface (with one isospin flavor) is occupied in these density ranges, each of which is thus a ferromagnetic spin- and valley-polarized state, akin to a quantum Hall ferromagnet, but driven by $D_\perp$-field.

\begin{figure*}
    \centering
    \includegraphics[width=6.75in]{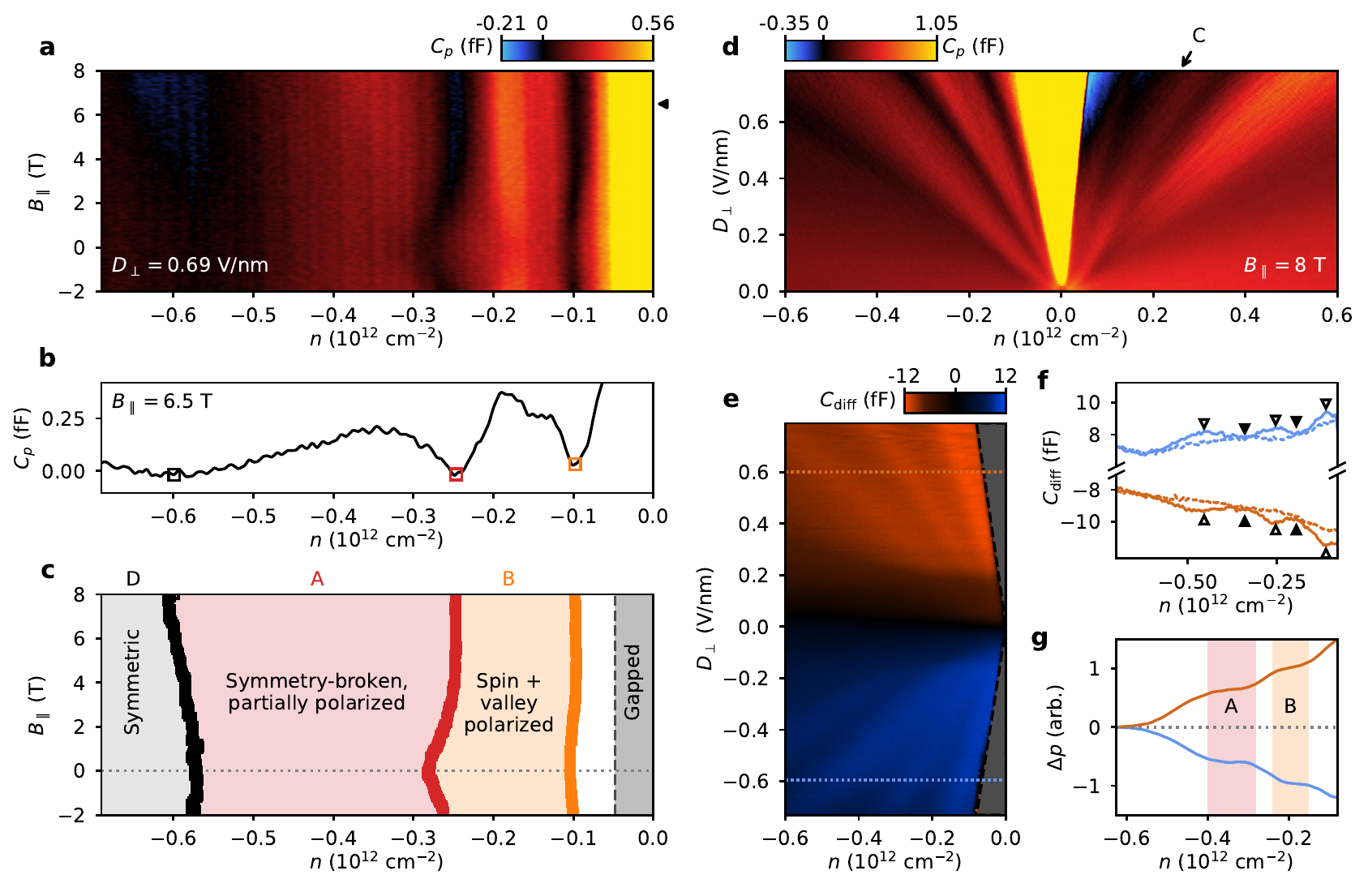}
    \caption{\figtitle{Phase transitions between spin and valley polarization.}
    \panela $B_\parallel$-field dependence of $C_p$ versus $n$ measured at fixed $D_\perp=\SI{0.69}{V/nm}$ ($B_\perp=0$).
    \panelb Line trace at $B_\parallel=\SI{6.5}{T}$ showing the series of minima corresponding to transitions between regions A, B, and D on the hold side of charge neutrality.
    \panelc Evolution of the $C_p$ minima with $B_\parallel$, which distinguishes phase boundaries with changing spin order from transitions that do not involve spin.
    \paneld $n$-$D_\perp$ map of $C_p$ measured at fixed $B_\parallel=\SI{8}{T}$, showing sharper boundaries between the symmetry-broken phases due to a trend toward increasingly negative compressibility between the phases with increasing $B_\parallel$, including a better-developed region C compared to $B=0$.
    \panele $n$-$D_\perp$ map and \panelf line traces of differential capacitance ($C_\text{diff}$, as described in main text) measured in Device~II at \SI{1.8}{K}, showing signatures of charge transfer between the layers near the boundaries (indicated by empty symbols) of each symmetry-broken phase (indicated by solid symbols). Comparison of \SI{1.8}{K} data in \panelf (solid) to \SI{40}{K} data (dashed) reveals enhancements in $C_\text{diff}$ at the phase boundaries.
    \panelg Excess layer polarization $\Delta p$ calculated by integrating the difference between high- and low-temperature differential capacitance, showing enhanced layer polarization in symmetry-broken states A and B.
    }
    \label{fig4}
\end{figure*}

To determine the precise nature of the symmetry-broken states, it is necessary to experimentally probe the flavor degrees of freedom.
Using the fact that in-plane magnetic fields couple preferentially to spin, we can follow the evolution of phase boundaries between the symmetry-broken states in $B_\parallel$ to reveal changes in spin ordering (\figlbl~\ref{fig4}\panela).
Taking a cut along the density direction at fixed $D_\perp=\SI{0.69}{V/nm}$, we track the positions of capacitance minima, taken to be the boundaries of each symmetry-broken phase, while sweeping a pure $B_\parallel$ field (\figlbl\ref{fig4}\panelb).
The resulting A--B phase boundary decreases in density as the magnitude of $B_\parallel$ is increased (\figlbl.~\ref{fig4}\panelc).
If the primary effect of $B_\parallel$ is to introduce a Zeeman splitting of the spin states in the system, we can interpret the reduction in density of the A--B transition point as resulting from the difference in Zeeman energy between the two phases with $B_\parallel$ \cite{zhou2021halfandquarter}.
From fixed field measurements (\figlbl~\ref{fig2}) and quantum oscillations (\figlbl~\ref{fig3}) we can attribute region B to a fully symmetry-broken one-fold ($1\times$)-degenerate state with full spin and valley polarization.
The change in the A--B boundary suggests that the spin order differs between the two phases, suggesting that region A is symmetry-broken, but without complete spin polarization.
The fully symmetric, $4\times$-degenerate D region is also spin-unpolarized; all four spin-valley flavors are equally occupied.
The D--A phase boundary also shifts slightly in the presence of a large in-plane field.
Together with the fact that the width of region A along the density axis (and thus the total degeneracy of the state) increases as $B_\parallel$ is applied, this suggests that region A may posses some degree of partial spin polarization.
Considering the two-fold features in region A observed in fixed $B_\perp$, weaker in comparison to region B, one possibility is that region A is only partially polarized, with some additional flavor occupation beyond the two leading flavors that result in the peak at $1/2$ in the Fourier analysis (curve A, \figlbl.~\ref{fig3}\panelc).
This interpretation is consistent with the observation that an additional, low-frequency peak emerges in region A below $1/4$, indicating that occupation is shared between at least two Fermi surfaces with differing $k$-space geometry.

In ABC trilayer graphene, a similar sequence of phases was observed for electrons and holes, however, in the latter case the $2\times$- and $1\times$-degenerate states were interrupted by intervening states with incomplete isospin polarization and annular Fermi surfaces \cite{zhou2021halfandquarter}.
Furthermore, the nature of a $2\times$-degenerate electron state in ABC trilayer graphene was shown to be spin polarized, in contrast to the $2\times$-degenerate hole state in bilayer graphene, likely owing to quantitative differences in the precise exchange couplings in each system.
One generic feature that appears to be shared between the cases of ABC trilayer and bilayer graphene is the lack of additional $3\times$-degenerate states.
Such states were predicted in a simple Stoner model for ABC trilayer graphene \cite{zhou2021halfandquarter}, but do not appear to manifest in either system.

Another effect of $B_\parallel$ on the system is that the boundaries between the symmetry-broken states are more clearly resolved (\figlbl.~\ref{fig4}\paneld) compared to $B=0$.
The visibility of each phase boundary is increased in the $n$-$D_\perp$ map due to decreasing $C_p$ minima, trending toward negative compressibility as $B_\parallel$ is increased (as in \figlbl.~\ref{fig4}\panelb).
Negative compressibility is an indication of electronic ordering in the system, typically emerging in proximity to phase transitions into such ordered states \cite{eisenstein1992negative}.
As such, this trend supports the interpretation of the sharp $C_p$ minima as occurring at or near phase boundaries between the symmetry-broken phases.
Notably, the boundaries of the $1\times$-degenerate electron state (region C) are more clearly resolved at $B_\parallel=\SI{8}{T}$ compared to $B=0$, lending further support to the assignment of the symmetry-broken state observed in Figs.~\ref{fig2} and \ref{fig3}.
The phase boundary of the $1\times$-degenerate state on the electron side (region C) for small values of $B_\parallel$ is not as clear as for holes, however, and therefore it is challenging to comment on the spin ordering based on the phase boundary alone.
Nevertheless, only one Fermi surface is filling in this state (\figlbl.~\ref{fig2}), suggesting that this is a spin- and valley-polarized state, similar to region B.
It is possible that other symmetry-broken phases appear for electrons at larger displacement fields or in a narrow density ranges bordering region C, but no clear intermediate phases emerge in our measurements.
Similarly, a narrow window of densities exists between the $1\times$-degenerate regions B and C and the band gap at large displacement fields for which we are unable to make a confident assignment of the degeneracy or isospin ordering.
We leave these regions to future investigation with even higher quality samples that display quantum oscillations at lower magnetic fields.

To further probe the character of the symmetry-broken states, we make use of layer sensitivity provided by the capacitance measurement.
By applying out-of-phase signals to the top and bottom gates of the sample while measuring the response on the bilayer, we null the contribution from electronic compressibility and instead probe the tendency of the system to polarize across the layers of the bilayer \cite{hunt2017direct}.
This measurement, the differential capacitance $C_\text{diff}$, is positive for one sign of the displacement field and negative for the other (\figlbl.\ref{fig4}\panele).
The measured $C_\text{diff}$ is related to layer polarization $p = n_1 - n_2$ (for layer densities $n_1$ and $n_2$) by integration, $p \propto \int C_\text{diff}\ dn$, and thus the sign indicates the direction of polarization, whereas the magnitude of $C_\text{diff}$ indicates the tendency of the system to fill one layer or the other as carriers are added.
The overall slope of $C_\text{diff}$ in \figlbl.~\ref{fig4}\panele-\panelg largely reflects the simple electrostatic effect of $D_\perp$, however the small oscillations observed on top of this background are related instead to the symmetry-broken states.
The regions indicated by filled symbols in \figlbl.~\ref{fig4}\panelf align with the symmetry-broken regions A and B measured in $C_p$, whereas the empty symbols denote the phase boundaries.
By comparing curves measured at \SI{1.8}{K} (solid) to ones measured at \SI{40}{K} (dashed) where the signature of the symmetry-broken states is quenched, we find that $C_\text{diff}$ is enhanced at the phase boundaries at low temperature.
That is, the magnitude of $C_\text{diff}$ is larger and suggests that charge may transfer more readily between the layers at the boundaries of each symmetry-broken phase.
Taking the difference of the high- and low-temperature data $\Delta C_\text{diff} = C_\text{diff}^{\SI{40}{K}} - C_\text{diff}^{\SI{1.8}{K}}$ and integrating along the density direction, we obtain the enhancement of the layer polarization due to the symmetry-broken phases, $\Delta p = \int \Delta C_\text{diff}\ dn$ (\figlbl~\ref{fig4}\panelg).
This excess polarization $\Delta p$ shows that the degree of layer polarization is enhanced within the symmetry-broken phases, suggesting some charge transfer occurs between the layers due to isospin polarization.

Exchange splitting driven by displacement-field-induced transitions is an exciting direction for further study in bilayer graphene.
There is an interesting parallel between the role of the $D_\perp$ field in driving the bilayer into isospin polarized states and quantum Hall ferromagnetism.
Whereas $B_\perp$ field can induce ferromagnetism by forming extremely flat Landau levels at odd integer fillings in any sufficiently clean 2D electronic system, in bilayer graphene the electric displacement field provides an alternative tuning parameter for generating flat bands.
Crucially, $D_\perp$ can drive the system into isospin polarized states at experimentally accessible $D_\perp$ fields at $B=0$.
This mechanism is reminiscent of, but distinct from the observation of correlated states in \moire systems where a periodic superlattice induces flat bands and large densities of states.
Here, no superlattice is required and, more importantly, the degree of band flatness is tunable by an external parameter \textit{in situ}.
This presents a distinct advantage over many \moire systems where the bandwidth is fixed by the angle of rotation or mismatch between constituent lattice periodicities.
For this reason, and due to the ease with which bilayer graphene can be prepared and measured experimentally relative to ABC trilayers and other graphene multilayers, bilayer graphene is a strong candidate for further investigation in the context of correlated physics.

While preparing this manuscript we became aware of related work \cite{zhou2021isospin} reporting the existence of superconductivity in bilayer graphene at lower temperatures in proximity to symmetry-broken states.

\section*{Methods}
\subsection{Device fabrication}

The devices in this study were each fabricated using a single boron nitride (BN) crystal cut into two pieces to ensure a symmetric gating environment for the bilayer graphene (BLG).
Each BN flake was etched into two using a reactive ion etching method.
We first picked up one of the two BN flakes with a graphite flake and put them onto a substrate with pre-patterned markers.
The stack was annealed at \SI{350}{\celsius} for at least 3~hours with \ce{H2}/Ar gas.
Then, the other half of the BN flake was used to pick up the BLG and transferred onto the BN/graphite stack.
Another graphite flake was then put on top in a separate transfer process.
The whole device structure was assembled using standard dry transfer techniques.
Graphite gates were deliberately employed to minimize charge disorder and ensure a high device quality.
Afterwards, electrical contact was defined by e-beam lithography and made through a top contact method using Cr/Pd/Au.  

For Device~I, the top and bottom BN flakes were aligned at \SI{0}{\degree}, while in Device~II, the two BN flakes were rotated by \SI{90}{\degree} with respect to one another.
In all measurements, no superlattice peaks were observed.
Therefore, it is unlikely that the BLG is aligned to any of the BN flakes.
We believe that the relative BN angles are not important for our reported observations as we also performed the same measurement in a purposely misaligned sample and the same phenomenon was consistently observed. 
All data in the main text were measured in Device~II unless otherwise specified.

\subsection{Capacitance measurements}

All capacitance data in the study were measured by balancing a cryogenic capacitance bridge circuit at a fixed point in parameter space and measuring the off-balance voltage via a low-temperature amplifier circuit.
Sinusoidal ac voltage signals were applied simultaneously to the device and a reference capacitor whose value was measured separately.
The phase and amplitude of the signal applied to the reference capacitor was varied in order to null the signal at the balance point of the bridge circuit.
As the capacitance of the device changes, an off-balance voltage develops at the balance point whose magnitude is proportional to the device capacitance.

The sample and reference capacitances are far smaller than parasitic capacitances of the cryostat cabling and are difficult to measure directly due to signal shunting.
A custom low-temperature amplifier is therefore fabricated and placed as close to the sample as possible in the cryostat (typically a few millimeters away).
This amplifier transforms the impedance of the bridge circuit and provides a small amount of voltage gain.
In penetration field capacitance measurements ($C_p$) the amplifier and bridge balance point was connected to the back gate of each device, whereas in differential capacitance measurements ($C_\text{diff}$) they were connected to the bilayer graphene.
Excitation signals in this study were $V_\text{ac} = \SI{5}{mV_{rms}}$ applied at a frequency of \SI{51.747}{kHz} either to the top gate (as in $C_p$ measurements) or to the top and bottom gates simultaneously (as in $C_\text{diff}$ measurements).
All measurements were performed between \SI{40}{mK} and \SI{100}{mK} unless otherwise specified.

Differential capacitance measurements were performed by tuning the amplitude of two out-of-phase signals applied to the top and bottom gates simultaneously to null the contribution of the electronic compressibility at the balance point.
Due to careful device design and the use of symmetric BN layers, the optimal ratio of top and bottom excitation amplitudes was exactly 1.
Parallel capacitance from unintentional geometric contributions was separately nulled using a signal applied to the reference capacitor in the same fashion as in $C_p$ measurements.

\setcitestyle{numbers}

\end{document}